\documentclass{article}
\usepackage{amssymb,graphicx,epsfig}

\font\tenbg=cmmib10 at 10pt
\def \rvecxi{{\hbox{\tenbg\char'030}}}
\def \rvecphi{{\hbox{\tenbg\char'036}}}
\def \rvecdelta {{\hbox {\tenbg\char'016}}}
\def \rvecepsilon {{\hbox {\tenbg\char'017}}}

\begin{document}

\title{Negative Energy and Angular Momentum Modes of Thin Accretion Disks}
\date{}
\author{L. Zhang\thanks{Pennsylvania State University, University Park, PA 16802; LXZ1@psu.edu} \and R.V.E. Lovelace\thanks{Department of Astronomy, Cornell University, Ithaca, NY 14853; RVL1@cornell.edu}}
\maketitle

\begin{abstract}

This work derives the linearized
equations of motion, the
Lagrangian density, the Hamiltonian
density, and the
canonical angular momentum density
for general perturbations [$\propto
\exp(im\phi)$ with $m=0,\pm 1,..$]
of a geometrically thin
self-gravitating, homentropic fluid disk
including the pressure.
The theory is applied
to ``eccentric,'' $m=\pm 1$ perturbations
of a geometrically thin Keplerian disk.
We find $m=1$ modes at
low frequencies relative to the Keplerian
frequency. Further, it shown that
these modes can have negative energy and
negative angular momentum.
The  radial
propagation of these low frequency $m=1$ modes  can
transport angular momentum away from the inner region of a disk
and thus increase the rate of mass accretion.
Depending on the radial boundary conditions there
can be discrete low-frequency, negative-energy, $m=1$
modes.

\bigskip

\noindent{key words: accretion,
accretion disks---instabilities---
galaxies: kinematics and dynamics}

\end{abstract}

\section{Introduction}

Problems of long standing interest
concern the linear modes and
instabilities of accretion disks
and the disks of spiral galaxies
(for example, Kato, Fukue, \&
Mineshige 1998; Binney \& Tremaine 1987).
Interest in
unstable modes in accretion disks has
been stimulated by recent observations of
quasi-periodic oscillations (QPOs) in
accreting neutron stars and black hole
candidates in binary systems
(see review by van der Klis 2000).
On a larger scale, recent observations
of disk galaxies reveal that about
$30\%$ of the disks are ``lopsided''
(Baldwin, Lynden-Bell, \& Sancisi 1980;
Rix \& Zaritsky 1995; Kornreich,
Haynes, \& Lovelace 1998).
In some cases the lopsidedness
may be dynamical in origin and
an indication of an unstable
one-armed trailing spiral wave
(Lovelace, Zhang, Kornreich, \& Haynes
1999; hereafter LZKH). Hybrid,
N-body/hydrodynamic simulations of
impulsively perturbed disk galaxies show
 long-lived trailing spiral features
(Zeltwanger {\it  et al.} 2000; Kornreich
{\it et al. }2002).

In previous work we derived equations
of motion, a Lagrangian,
Hamiltonian, and canonical
angular momentum for the eccentric
perturbations
[$\propto \exp(\pm i\phi)$]
of a thin self-gravitating
disk of a galaxy consisting of a large
number of rings (LZKH).
This work utilized a Lagrangian description
of the perturbation.
The present work generalizes LZKH
by deriving equations of motion,
a Lagrangian density, Hamiltonian
density, and
canonical angular momentum density
for general perturbations [$\propto
\exp(im\phi)$ with $m=0,\pm 1,..$]
of a self-gravitating fluid disk
including the pressure.
The theory allows the calculation
of the energy and angular momentum of the
different modes and identification
of modes with {\it negative energy} and/or
{\it negative angular momentum}.
The negative energy modes
can become unstable in the
presence of dissipation.
Negative energy modes are important
in plasma physics (Coppi, Rosenbluth,
\& Sudan 1969),
in Rossby r-mode oscillations
of rapidly rotating neutron stars
which are made unstable by gravitational radiation
(Andersson 1998; Schenk {\it et al.} 2002), and for
short wavelength modes
in the shearing sheet
model of modes of disks
(Goodman, Narayan, \& Goldreich 1987).
We apply our theory to the ``eccentric,'' $m=1$
perturbations of a geometrically thin Keplerian disk.
We find $m=1$ modes with frequencies much
less than the Keplerian frequency.
These modes  can have negative energy and
negative angular momentum.
The radial
propagation of such a mode  may act to remove
angular momentum from the inner region of a disk.
Earlier, Nowak and Wagoner (1991, 1992) analyzed
the modes of  non--self-gravitating
disks using a Lagrangian description of the
perturbation.
The important role of $m=1$ modes for
accretion disk angular momentum transport has
been discussed by Fridman {\it et al.} (2003).
Adams, Ruden, and Shu (1989)
studied the growth of global
gravitational instabilities in star/disk systems
and found modes which grow on a
nearly dynamical time scale.
  Shu et al. (1990)
presented an analytical description of
the corresponding modal mechanisms.
    Tremaine (2001) analyzed
slow modes using softened gravity and
found these modes to be stable.

In \S 2 we derive the linearized
equations of motion and obtain
the Lagrangian density. From
this the Hamiltonian density and
canonical angular momentum density
are derived in the usual way.
In \S 3 we introduce a
representation for the perturbation
amplitudes in terms of complex
quantities.
In \S 4 the WKBJ approximation
is introduced for the radial
dependence of the perturbations,
and the results of the present
approach are contrasted with the standard
analysis.
In \S 5 we consider the inner
region of a low mass disk around a
black hole and show that
one of the $m=1$ modes has a very
low frequency (compared with the
rotational frequency).
This mode is shown to
have negative energy which means
that it may be made unstable by
dissipation.
In \S 6 we give conclusions
of this work.

\section{Theory}

We use an inertial
cylindrical $(r,\phi,z)$ coordinate
system and assume that the
disk is geometrically thin
with half-thickness $h \ll r$.
Further, we consider conditions
where the accretion speed $-v_r \ll v_\phi$.
We use a
Lagrangian representation
for the perturbation
as developed by Frieman and Rotenberg (1960)
and Lynden-Bell and Ostriker (1967) and
as applied to a system of rings
representing a self-gravitating disk
(LZKH; Lovelace 1998).
The position vector ${\bf r}$ of a fluid
element which at $t=0$
was at ${\bf r}_0$ is given by
\begin{equation}
{\bf r}(t) = {\bf r}_0(t) +\rvecxi({\bf r}_0,t)~.
\end{equation}
That is, ${\bf r}_0(t)$ is the unperturbed and
${\bf r}(t)$ the perturbed orbit of a fluid element
as sketched in Figure 1.

Further, the perturbations
are assumed to consist
of small in-plane displacements
of the disk matter,
\begin{equation}
\rvecxi =\xi_r \hat{{\bf r}}+
\xi_\phi\hat{\rvecphi~}~.
\end{equation}
The
$\xi_r$ and $\xi_\phi$
are proportional to
$\exp(im\phi)$ with $m=0,1,2,...$
the azimuthal mode number.
(The negative values of $m$ are
redundant.)

\begin{figure}\begin{center}
\includegraphics[width=5in]{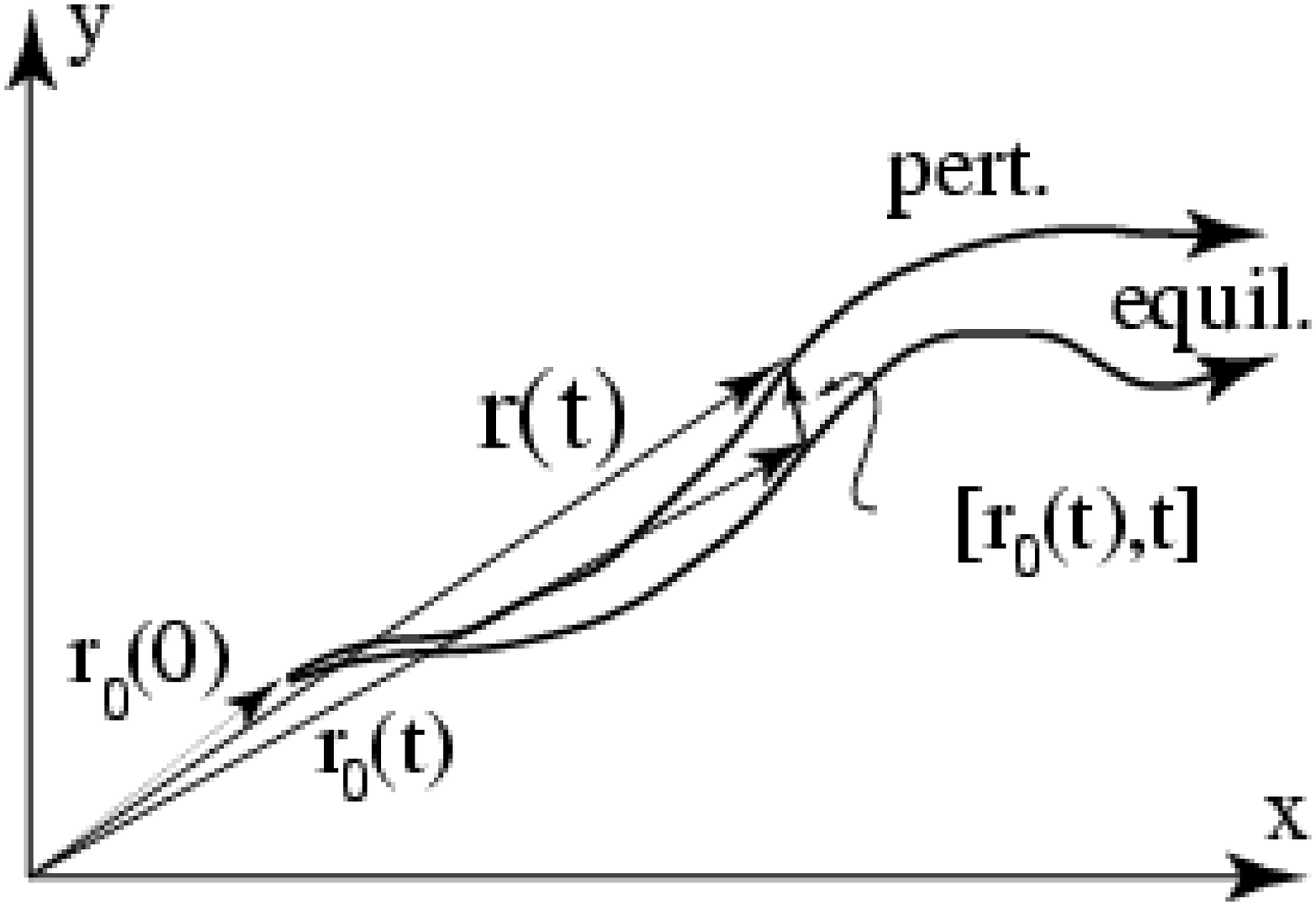}
\caption{Diagram illustrating
equation (1), where ``equil.'' indicates
the equilibrium path of a fluid element
and ``pert.'' indicates the perturbed
path. See Frieman \& Rotenberg (1960).}
\end{center}
\end{figure}


 From equation (1), we have
${\bf v}({\bf r},t)={\bf v}_0({\bf r}_0,t)+
\partial \rvecxi/\partial t +
({\bf v} \cdot {\bf \nabla})
\rvecxi$.
The Eulerian velocity perturbation is
$\delta {\bf v}({\bf r},t) \equiv {\bf v}({\bf r},t)
-{\bf v}_0({\bf r},t)$.
Therefore,
\begin{equation}
\delta{\bf v}({\bf r},t) =
{\partial \rvecxi \over \partial t}
+\left({\bf v} \cdot {\bf \nabla}\right) \rvecxi
-\left(\rvecxi \cdot {\bf \nabla} \right) {\bf v}~.
\end{equation}
The components of this equation are
\begin{eqnarray}
\delta v_r &=& {\cal D} ~\!\xi_r~, \nonumber \\
\delta v_\phi &=& {\cal D} ~\!\xi_\phi -
r \Omega^\prime~\! \xi_r~,
\end{eqnarray}
where
$$
{\cal D} \equiv {\partial \over \partial t}
+\Omega(r){\partial \over \partial \phi}~,
$$
with $\Omega(r)$ the angular rotation rate
of the equilibrium disk and
$\Omega^\prime \equiv d \Omega/ dr$.

The equation of motion is
\begin{equation}
{d~ \delta {\bf v} \over dt}
=\delta {\bf F}= -{1\over \Sigma}{\bf \nabla } \delta P+
{\delta \Sigma \over \Sigma^2}{\bf \nabla}P
-{\bf \nabla}\delta \Phi~,
\end{equation}
where $\delta {\bf F}$ is the perturbation
force per unit mass,
$\delta \Phi$ is
the perturbation of the gravitational
potential, $\Sigma$ is the disk's surface
mass density, and $P$ is the vertically
integrated pressure in the disk.
For simplicity we consider a
homentropic disk where
$P ={\rm const}~ \Sigma^\Gamma$.
Under this condition,
$\delta {\bf F}=
-{\bf \nabla } (\delta P / \Sigma)
-{\bf \nabla}\delta \Phi.
$

We have
\begin{equation}
{d~ \delta{\bf v} \over dt}=
{\partial ~\delta {\bf v} \over \partial t}
+\left({\bf v}\cdot {\bf \nabla} \right) \delta {\bf v}
+\left(\delta {\bf v} \cdot {\bf \nabla} \right) {\bf v}~.
\end{equation}
The components of equation (6) give
\begin{eqnarray}
\left({d~ \delta {\bf v} \over dt }\right)_r &=&
{\cal D} \delta v_r -2 \Omega \delta v_\phi~, \nonumber \\
&=& ({\cal D}^2 +2\Omega r \Omega^\prime)~\! \xi_r
-2\Omega{\cal D}~\!\xi_\phi~, \nonumber\\
\left({d~ \delta {\bf v} \over dt} \right)_\phi
&=& {\cal D} \delta v_\phi
+(\kappa_r^2/2\Omega) \delta v_r~, \nonumber \\
&=& {\cal D}^2 ~\!\xi_\phi +2\Omega{\cal D} ~\!\xi_r~,
\end{eqnarray}
where $\kappa_r^2 \equiv (1/r^3)d(r^4\Omega^2)/dr$
is the radial
epicyclic frequency (squared).
For a low mass, geometrically thin Keplerian disk,
$\kappa_r^2 = \Omega^2 +{\cal O}(c_s^2/r^2) \approx \Omega^2$.
More generally, for $\Omega \propto 1/r^q$,
$\kappa_r^2= 2(2-q)\Omega^2$.
For a Keplerian
disk, $q=3/2$; for a spiral galaxy with a flat
rotation curve, $q=1$; and for a rigidly rotating disk,
for example, the inner region of a spiral galaxy,
$q=0$.

For $\delta{\bf F}=0$, equations (7) with $\xi_r,~\xi_\phi
\propto \exp(im\phi -i\omega t)$ can be solved to give four
roots: two have $\omega-m\Omega=0$ and $\xi_r=0,~\xi_\phi\neq 0$,
and two have $\omega-m\Omega = \pm \kappa_r$.
For the second pair of modes, ${\cal R}e(\xi_r)=
a\cos(\kappa_rt)$ and ${\cal R}e(\xi_\phi)
=-(2\Omega/\kappa)a \sin(\kappa_rt)$,
where $a=$const. and ${\cal R}e$ indicates
the real part; the fluid particle
moves along a retrograde ellipse which
is elongated in the azimuthal direction
since $\kappa_r < 2\Omega$ for most conditions
(Binney \& Tremaine 1987, p. 132).
For a rigidly rotating disk, $2\Omega/\kappa_r=1$
so that the ellipse is a circle.

The perturbation of the surface mass
density of the disk obeys
\begin{eqnarray*}
{\partial~\delta \Sigma \over \partial t}
+{\bf \nabla}\cdot \left(\Sigma ~\delta {\bf v}
+\delta \Sigma ~{\bf v} \right)=0~,
\end{eqnarray*}
where $\Sigma(r)$ is the surface density of the
equilibrium disk.
Because ${\bf \nabla} \cdot
(\Sigma~{\bf v}) = 0$, this equation implies
\begin{equation}
\delta \Sigma = - {\nabla}\cdot
\left(\Sigma~\rvecxi\right)~.
\end{equation}
The perturbation of the gravitational potential
is given by
\begin{equation}
\delta \Phi({\bf r},t) = - G \int d^2 r^\prime~
{\delta \Sigma({\bf r}^\prime,t)
\over \left| {\bf r} - {\bf r^\prime} \right|}~,
\end{equation}
where the integration is over the
surface area of the disk, and $G$ is
the gravitational constant.

\subsection{Perturbation Amplitudes}

We can write in general
\begin{eqnarray}
\xi_{r} &=& ~~\epsilon_{x}\cos(m\phi) +
\epsilon_{y}\sin(m\phi)~,\nonumber \\
\xi_{\phi}&=& -\delta_{x}\sin(m\phi) +
\delta_{y}\cos(m\phi)~,
\end{eqnarray}
where
$\epsilon_{x,y}(r,t)$ and $\delta_{x,y}(r,t)$ are
the {\it displacement amplitudes} for the `ring'
of the disk at radius $r$.
For $m=1$, $\epsilon_{x,y}$
represents the shift of the ring's center, and
$\delta_{x,y}$ represents in general both
the shift of the ring's center {\it and} the
azimuthal displacement of the ring matter
(see LZKH).

\subsection{Equations of Motion}

From equations (5) - (7) and (10), we
obtain the disk equations of motion,
\begin{eqnarray}
\ddot{\epsilon}_x+2m\Omega\dot{\epsilon}_y
-m^2\tilde{\Omega}^2\epsilon_x
-2\Omega (\dot{\delta}_y -m\Omega \delta_x)\!\!& =&
{\delta F}_r^C,
\nonumber \\
\ddot{\epsilon}_y-2m\Omega\dot{\epsilon}_x
-m^2\tilde{\Omega}^2\epsilon_y
+2\Omega (\dot{\delta}_x +m\Omega \delta_y)\!\!& =&
{\delta F}_r^S,
\nonumber \\
\ddot{\delta}_x+2m\Omega\dot{\delta}_y -m^2\Omega^2\delta_x
-2\Omega(\dot{\epsilon}_y -m\Omega {\epsilon}_x)\!\!& =&
\!\!\!-{\delta F}_\phi^S,
\nonumber \\
\ddot{\delta}_y-2m\Omega\dot{\delta}_x -m^2\Omega^2\delta_y
+2\Omega(\dot{\epsilon}_x +m\Omega {\epsilon}_y) \!\!&=&
{\delta F}_\phi^C,
\nonumber\\
\end{eqnarray}
where $\dot{\epsilon}_x \equiv
\partial \epsilon_x(r,t)/\partial t$, etc.
Here,
$$
\tilde{\Omega}^2 \equiv
\Omega^2 -2\Omega r\Omega^\prime/m^2~,
$$ and
$$
{\delta F}_\alpha^{C,S}
\equiv \oint {d\phi \over \pi}~
\big[\cos( m\phi),~\sin( m\phi)\big]
\delta F_\alpha
~,
$$
with the $S, C$ superscripts
denoting the sine and
cosine components, and
$\alpha = r,~\phi$.

The force per unit mass
\begin{equation}
\delta F_\alpha =
-{\bf \nabla }{\delta P \over \Sigma}
-{\bf \nabla}\delta \Phi=
{\bf \nabla}\left[{c_s^2{\bf \nabla}
\cdot(\Sigma \rvecxi)\over\Sigma}\right]
-{\bf \nabla}\delta\Phi~,
\end{equation}
with
$\alpha = r,~\phi$, and
$c_s^2 = d P/d\Sigma$ is the effective
sound speed (squared).
The pressure force on the
right-hand side of equation (12) gives
\begin{eqnarray}
{\delta F}_r^{Cp} &=&
D\big[c_s^2 D_\star(\Sigma
\epsilon_x)/\Sigma\big]-D(c_s^2m\delta_x/r)~,
\nonumber \\
{\delta F}_r^{Sp} &=&
D\big[c_s^2 D_\star(\Sigma
\epsilon_y)/\Sigma\big]-D(c_s^2m\delta_y/r)~,
\nonumber \\
-{\delta F}_\phi^{Sp} &=&
[mc_s^2/(r\Sigma)] D_\star(\Sigma \epsilon_x)
-c_s^2m^2\delta_x/r^2~,
\nonumber \\
{\delta F}_\phi^{Cp} &=&
[mc_s^2/(r\Sigma)] D_\star(\Sigma \epsilon_y)
-c_s^2m^2\delta_y/r^2,
\end{eqnarray}
where $D(..) \equiv \partial(..)/\partial r$ and $D_\star(..)
\equiv (1/r)\partial (r ~..)/\partial r$.

The gravitational force gives
\begin{eqnarray}
\delta F_r^{Cg} & =&
-{\partial \over \partial r}
G\int{ d^2r^\prime ~{\cal K}_m R_x^\prime}~,
\nonumber \\
\delta F_r^{Sg} & =&
-{\partial \over \partial r}
G\int{ d^2r^\prime ~{\cal K}_m R_y^\prime}
\nonumber \\
-\delta F_\phi^{Sg} &=&
-{m \over r}
G\int{ d^2r^\prime ~{\cal K}_m R_x^\prime}~,
\nonumber\\
\delta F_\phi^{Cg} &=&
-{m \over r}
G\int{d^2r^\prime~{\cal K}_m R_y^\prime}~,
\end{eqnarray}
where $d^2r^\prime =2 \pi r^\prime dr^\prime$,
$R^\prime_j=R_j(r^\prime,t)$,
\begin{equation}
R_j(r) \equiv D_\star(\Sigma \epsilon_j) -m \Sigma \delta_j/r
\end{equation}
for $j=x,y$, and
\begin{equation}
{\cal K}_m(r,r^\prime) \equiv
{1\over 2\pi} \oint {d\Psi \cos(m\Psi)
\over \sqrt{r^2+(r^\prime)^2 -2rr^\prime \cos(\Psi)}}~.
\end{equation}
We can express this as
$$
{\cal K}_m(r,r^\prime)={ (-2)^m \over (2m-1)!!}
{ P^m_{-1/2}(z) \over \sqrt{|r^2-(r^\prime)^2|}}~,
$$
where $z\equiv[r^2+(r^\prime)^2]/|r^2-(r^\prime)^2|$,
$(-1)!!=1$,
and $P^m_\nu$ is the usual
associated Legendre function.

\subsection{Lagrangian}

The Lagrangian which gives the
above equations of motion is
\begin{equation}
L = \int d^2r~{\cal L}=2\pi\int rdr~{\cal L}~,
\end{equation}
where ${\cal L}={\cal L}_0+{\cal L}_1
+{\cal L}_2+{\cal L}_3$ is the Lagrangian
density with
$$
{\cal L}_0 ={\Sigma \over 2}
\left[\dot{\rvecepsilon~}^2
+ \dot{\rvecdelta~}^2
+m^2\tilde{\Omega}^2 \rvecepsilon^2
+m^2\Omega^2 \rvecdelta^2
-4m\Omega^2 \rvecepsilon \cdot
\rvecdelta \right],
$$
$$
{\cal L}_1 =- \Sigma \Omega
\left[m{\rvecepsilon \times
\dot{\rvecepsilon~}}
+m\rvecdelta \times \dot{\rvecdelta~}
-\rvecdelta \times \dot{\rvecepsilon~}
-\rvecepsilon \times \dot{\rvecdelta~}
\right]\cdot~ \hat{\bf z}~,\quad
$$
\begin{eqnarray}
{\cal L}_2 &=& -{c_s^2
\over
2\Sigma} \big[{\bf \nabla}\cdot
(\Sigma \rvecxi)\big]^2 \nonumber
\\ &=&
-{1 \over 2}\Sigma c_s^2\left({\bf Q}
-{m \rvecdelta
\over r}\right)^2~,\nonumber
\end{eqnarray}
\begin{equation}
{\cal L}_3={G\over 2}\Sigma
\left({\bf Q}-{m \rvecdelta \over
r}\right)
\cdot
\int d^2r^\prime {\cal K}_m(r,r^\prime)
\Sigma^\prime \left({\bf Q}^\prime -
{m \rvecdelta^\prime \over r^\prime}\right)~.
\end{equation}
Here,
$\rvecepsilon \equiv \epsilon_x\hat{\bf x}
+\epsilon_y \hat{\bf y}$,
$\rvecdelta \equiv \delta_x\hat{\bf x}
+\delta_y \hat{\bf y}$,
$Q_j(r) \equiv D_\star(\Sigma \epsilon_j)/\Sigma$ ($j=x,y$),
${\bf Q} \equiv Q_x\hat{\bf x} + Q_y\hat{\bf y}$, and
$\Sigma^\prime=\Sigma(r^\prime)$.

The Lagrangian density is ${\cal L}=
{\cal L}[\rvecepsilon, \rvecdelta,
\dot{\rvecepsilon},
\dot{\rvecdelta}, {\bf Q}].$
In terms of functional
derivatives, the
equations of motion are
\begin{eqnarray}
{\partial \over \partial t}~{\delta { L} ~\over \delta
\dot{\epsilon~}\!\!_j}&=& {\delta {L} ~\over \delta
\epsilon_j}-
\Sigma {\partial\over \partial r}\left({1\over \Sigma}
{\delta {L} \over \delta Q_j }\right)~,
\nonumber\\
{\partial \over \partial t} ~{\delta { L}~\over
\delta \dot{\delta~}\!\!_j}&=&
{\delta { L} \over \delta \delta_j}~,
\end{eqnarray}
(see for example Goldstein 1950, ch. 11).
For the contribution from ${\cal L}_3$,
$\delta { L}_3/\delta {\bf Q}
=G \Sigma
\int d^2 r^\prime {\cal K}_m~\Sigma^\prime
({\bf Q}^\prime - m\rvecdelta^\prime/r^\prime)$
and $\delta { L}_3 /\delta \rvecdelta
=-G\Sigma(m/r) \int d^2 r^\prime {\cal K}_m~\Sigma^\prime
({\bf Q}^\prime - m\rvecdelta^\prime/r^\prime)$.

\subsection{Hamiltonian}

The Hamiltonian density is
\begin{equation}
{\cal H} = \dot{\rvecepsilon~}\cdot{\partial {\cal L}
\over \partial \dot{\rvecepsilon~}} + \dot{\rvecdelta~}\cdot
{\partial {\cal L} \over \partial \dot{\rvecdelta~}}
- {\cal L}~.
\end{equation}
The time derivative of this equation gives
\begin{equation}
{\partial {\cal H} \over \partial t}
+ {1 \over r} {\partial \over \partial r}
\left(r~ {\cal F}_{Er}\right)=0~, \quad
{\cal F}_{Er} \equiv \dot{\rvecepsilon~}
\cdot{\partial {\cal L} \over \partial{\bf Q}}~,
\end{equation}
where ${\cal F}_{Er}$ is the radial
energy flux-density.
The explicit Hamiltonian density is
\begin{eqnarray}
{\cal H}\!&=& \!{\Sigma \over 2}
\left(\dot{\rvecepsilon~}^2 \!\!+\!\!
\dot{\rvecdelta~}^2 \!\!-\!
m^2\tilde{\Omega}^2\rvecepsilon^2 \!-\!
m^2\Omega^2 \rvecdelta^2 \!+\!
4m\Omega^2 \rvecepsilon \cdot \rvecdelta
\right) \nonumber \\
\!&+&\!{1 \over 2}\Sigma c_s^2\left({\bf Q}
-{m \rvecdelta
\over r}\right)^2 \nonumber\\
\!&-&\!{G\over 2}
\Sigma \left({\bf Q}-{m\rvecdelta
\over r}\right)\!\!\cdot\!\!\int
d^2r^\prime~{\cal K}_m
\Sigma^\prime
\left({\bf Q}^\prime -{m \rvecdelta^\prime
\over r^\prime}\right)
\end{eqnarray}
in terms of the Lagrangian variables.
Assuming vanishing energy fluxes
through the inner and outer radii of the
disk, we have
\begin{equation}
H \equiv \int d^2r~{\cal H}~,\quad {\rm with}
\quad {d H \over dt}=0~,
\end{equation}
where $H=E$ is the energy of the perturbation.

\subsection{Canonical Angular Momentum}

We can make a canonical transformation from
rectangular to polar coordinates,
$(\epsilon_x,\epsilon_y)\rightarrow
(\epsilon, \varphi)$,
$(\delta_x,\delta_y) \rightarrow (\delta, \psi)$
to obtain the Lagrangian as ${\cal L} =
{\cal L}(\epsilon,\delta,\varphi,\psi,
\dot{\epsilon},\dot{\delta},\dot{\varphi},\dot{\psi},
D_\star(\Sigma \epsilon)/\Sigma,
\partial \varphi/\partial r)$.
Note for example that $\dot{\rvecepsilon~}^2 \rightarrow
\dot{\epsilon}^2+\epsilon^2\dot{\varphi}^2$.
We then find that ${\cal L}$ is invariant
under the simultaneous changes
$\varphi \rightarrow \varphi +\theta$,
$\psi \rightarrow \psi + \theta$.
Therefore, it is useful to make
a further canonical transformation,
$[\varphi, \psi]\rightarrow [\alpha=(\varphi+\psi)/2,
\beta = \varphi -\psi]$, which gives
$\partial {\cal L}/\partial \alpha =0$.
Defining the canonical angular momentum
density as
$${\cal P}_\phi \equiv {\partial {\cal L}
\over ~\partial \dot{\alpha}}~,
$$
we obtain
\begin{equation}
{\partial {\cal P}_\phi \over \partial t}
+{1\over r}{\partial \over \partial r}
\left(r~{\cal F}_{{\cal P}r}\right)
=0~,\quad {\cal F}_{{\cal P}r} \equiv
{ \partial {\cal L} \over \partial (\partial
\alpha/\partial r)}~,
\end{equation}
where ${\cal F}_{{\cal P}r}$ is the
radial flux-density of angular momentum.
Assuming vanishing angular momentum
fluxes through the inner and outer
radii of the disk, we have
\begin{equation}
P_\phi =\int d^2r~ {\cal P}_\phi~, \quad {\rm with}\quad
{d P_\phi \over dt}=0~,
\end{equation}
where $P_\phi$ is the canonical angular
momentum of the perturbation.
From equation (20) we find
\begin{equation}
{\cal P}_\phi = \Sigma
\bigg[\epsilon^2(\dot{\varphi}-m\Omega)
+ \delta^2(\dot{\psi}-m\Omega)
+2\Omega\epsilon \delta \cos(\varphi-\psi)\bigg]~,
\end{equation}
where the last term can also be
written as $2 \Sigma \Omega \rvecepsilon \cdot
\rvecdelta$.

\section{Complex Displacement Amplitudes}

It is useful for later work to introduce
the complex displacement amplitudes,
\begin{eqnarray}
{\cal E} \equiv \epsilon_x - i \epsilon_y
= \epsilon(r,t) \exp \big[-i \varphi(r,t)\big]~, \nonumber \\
\Delta \equiv \delta_x - i \delta_y
= \delta(r,t) \exp\big[-i \psi(r,t)\big]~,
\end{eqnarray}
(LZKH).
Here, $\epsilon$ and $\delta$
are clearly non-negative, and the angles
$\varphi$ and $\psi$ are the
line-of-nodes angles for the ${\cal E}$
and $\Delta$ components of the
perturbation.
In terms of these complex amplitudes,
equations (11) can be combined to
give
$$
\ddot{\cal E}+2im\Omega \dot{\cal E}-
m^2 \tilde{\Omega}^2{\cal E}
-2i\Omega \dot{\Delta}+2m\Omega^2 \Delta=
\delta F_{\cal E}~,\quad~~
$$
\begin{equation}
\ddot{\Delta}+2im\Omega \dot{\Delta}-
m^2 {\Omega}^2{\Delta}
-2i\Omega \dot{\cal E}+2m \Omega^2{\cal E}=
\delta F_\Delta~,
\end{equation}
where $\delta F_{\cal E} \equiv
{\delta F}_r^C
-i{\delta F}_r^S$ and $\delta F_\Delta
\equiv
-\delta F_\phi^S
-i\delta F_\phi^C$.

\section{WKBJ Approximation}

For perturbations with short radial
wavelengths $\lambda_r = 2\pi/|k_r|$,
the WKBJ approximation is useful.
That is,
for $h < \lambda_r/(2\pi) < r$,
we can take
${\cal E},~\Delta \sim
\exp[i\int^r dr^\prime k_r(r^\prime)]$
in equation (18).
The radial derivatives can be approximated
as for example $D_\star(\Sigma {\cal E})\approx
ik_r \Sigma {\cal E}$.
The azimuthal wavenumber $k_\phi \equiv m/r$
is assumed to be less than or of the order of $|k_r|$
which corresponds to $0 \leq m < r/h$.
Note that the radial wavenumber $k_r$
may be $>0$ or $<0$. For $m \geq 1$, $k_r >0$
corresponds to a trailing spiral wave and
$k_r <0$ to a leading spiral wave.

With the WKBJ approximation,
the pressure force becomes
\begin{eqnarray}
\delta F_{\cal E}^p &=& -k_r^2 c_s^2 {\cal E}
-ik_r k_\phi c_s^2 \Delta~,\nonumber \\
\delta F_\Delta^p &=& -k_\phi^2 c_s^2 \Delta
+ik_r k_\phi c_s^2 {\cal E}~.
\end{eqnarray}
The gravitational force is
\begin{eqnarray}
\delta F_{\cal E}^g &=&
2\pi G \Sigma(k_r^2{\cal E} +i k_rk_\phi \Delta)/k~,
\nonumber \\
\delta F_\Delta^g &=& 2\pi G\Sigma
(k_\phi^2 \Delta -ik_r k_\phi {\cal E})/k~,
\end{eqnarray}
where $k \equiv (k_r^2 +k_\phi^2)^{1/2}$.

In terms of ${\cal E}$ and $\Delta$, the
Hamiltonian density in the WKBJ approximation
becomes
\begin{eqnarray}
{\cal H}\!\!&=&\!\!{\Sigma \over 2}\bigg(
\big|{\dot{\cal E}}\big|^2 +\big|{\dot{\Delta}}\big|^2
-(m^2\Omega^2-2\Omega r \Omega^\prime)\big|{\cal E}\big|^2
\nonumber \\
\!\!&-&\!\!m^2\Omega^2\big|\Delta\big|^2 + 4m\Omega^2 {\cal R}e
({\cal E} \Delta^\star) \bigg)
\nonumber\\
\!\!&+&\!\!{\Sigma u^2 \over 2}
\left|{ik_r {\cal E}}
- k_\phi \Delta \right|^2~,
\end{eqnarray}
where $u^2 \equiv c_s^2 -2\pi G \Sigma/k$,
${\cal R}e(..)$ indicates the real part,
and $(..)^\star$ the complex conjugate.
The radial energy flux-density
is
\begin{eqnarray}
{\cal F}_{Er}\!\!&=
&\!\! -\Sigma~ u^2 ~\dot{\rvecepsilon~}
\cdot ({\bf Q} -k_\phi {\rvecdelta})
\nonumber \\
\!\!&=&\!\! - \Sigma~u^2~{\cal R}e
\big[\dot{\cal E}^\star \left(
{ik_r {\cal E} } - k_\phi
\Delta \right) \big]~,
\end{eqnarray}
from equation (23).

Similarly, the canonical angular momentum
density can be written as
\begin{eqnarray}
{\cal P}_\phi =\Sigma \bigg\{ {\cal I}m\big[{\cal E}
(\dot{\cal E}^\star -im\Omega {\cal E}^\star)\big]
\!\!&+&\!\!{\cal I}m\big[\Delta (\dot{\Delta}^\star
-im\Omega \Delta^\star)\big] \nonumber\\
\!\!&+&\!\!2\Omega {\cal R}e({\cal E} \Delta^\star)\bigg\}~,
\end{eqnarray}
where ${\cal I}m(..)$ indicates the imaginary part.
The radial flux-density of
canonical angular momentum is
\begin{equation}
{\cal F}_{{\cal P}r}=\Sigma~u^2~
{\cal I}m \big[ {\cal E}^\star \left(
{ik_r {\cal E}}-
k_\phi\Delta \right)
\big]~,
\end{equation}
from equation (26).

\subsection{Mode Frequencies}

Using equations (28) and
taking ${\cal E},~\Delta \sim \exp(-i\omega t)$,
we obtain
\begin{equation}
\left[
\begin{array}{cc}
A_{11}&A_{12}\\
A_{21}&A_{22}
\end{array}
\right]
\left[
\begin{array}{c}
{\cal E} \\ \Delta
\end{array}
\right]=0~,
\end{equation}
where
\begin{eqnarray}
A_{11}&=&-\tilde{\omega}^2+2\Omega r \Omega^\prime
+(k_rc_s)^2-2\pi G\Sigma k_r^2/k~,
\nonumber \\
A_{12}&=&-2\Omega\tilde{\omega}
+ik_rk_\phi c_s^2 - 2\pi i G\Sigma k_r k_\phi/k
\nonumber \\
A_{21}&=&-2\Omega \tilde{\omega}
-ik_rk_\phi c_s^2+2\pi i  G \Sigma k_r k_\phi /k~,
\nonumber \\
A_{22}&=& -\tilde{\omega}^2
+ (k_\phi c_s)^2 - 2 \pi G \Sigma k_\phi^2/k~,
\end{eqnarray}
where $\tilde{\omega}(r)\equiv \omega - m \Omega(r)$ is
the wave frequency in the reference frame comoving
with the disk at $r$.

The dispersion relation implied by equation (35),
$\det(A)=0$, gives
$$
{\omega}_{1,2,3,4}=m\Omega ~\pm_A
$$
\begin{equation}
\left\{{1\over 2}\left[
\kappa_r^2+{\bf k}^2u^2 \pm_B
\sqrt{(\kappa_r^2+{\bf k}^2
u^2)^2 -8k_\phi^2 u^2\Omega r
\Omega^\prime}~\right]\right\}^{1 \over 2},
\end{equation}
where
the $(A,~B)$ subscripts on the $\pm$ signs
indicate that they are independent, and ${\bf
k}^2 \equiv k_r^2+k_\phi^2$. We adopt the
convention that $\omega_\alpha(A,B)=
\omega_1(+,~+),$ $~\omega_2(+,~-),
~\omega_3(-,~+),~{\rm and}~ \omega_4(-,~-)$.
The frequencies are invariant under
$k_r \rightarrow -k_r$ owing to the
time reversibility of the basic equations;
trailing spiral waves ($k_r>0$) are equivalent
to leading waves ($k_r<0$).

The mode frequencies found here
differ in some respects from the
standard analysis owing to our
different application of the WKBJ
approximation.
In the standard approach, one takes, for example,
${\bf \nabla}\cdot \delta {\bf v} \approx i{\bf k}\cdot
\delta {\bf v}$
(Binney \& Tremaine 1987, ch. 6;
Kato {\it et al.} 1998, ch. 13),
whereas we take
${\bf \nabla} \cdot \rvecxi \approx i {\bf k}
\cdot \rvecxi$.
Note that the standard approach does not
lead to equations of motion derivable
from a Lagrangian.

For axisymmetric $k_\phi=m/r=0$ perturbations,
equation (37) gives two zero frequency modes
and the two modes with
\begin{equation}
\omega = \pm \bigg(\kappa_r^2 +k_r^2 c_s^2 -
2\pi G \Sigma |k_r|\bigg)^{1/2}~,
\end{equation}
which is the same as
the classical result of
Toomre (1964) and Safronov (1960).
As a function of $k_r$, $\omega^2$ has
a minimum at $k_{TS}=\pi G \Sigma/c_s^2$
at which point $\omega^2(k_{r0})=
\kappa_r^2-(\pi G \Sigma/c_s)^2$.
If this minimum value is less than
zero, which occurs for
$\Sigma >\Sigma_{TS}
\equiv \kappa_r c_s/(\pi G)$, then
one of the $\omega$ values
is positive imaginary corresponding to a
linear Jeans-type instability
(Safronov 1960; Toomre 1964).
With $Q \equiv \Sigma_{TS}/\Sigma$,
there is instability for $Q<1$
(Toomre 1964).

For tightly wrapped spiral waves
($k_r^2 \gg k_\phi^2$), the two modes
with $\pm_B=+1$ simplify to
$\omega=m\Omega \pm \big[\kappa_r^2+k_r^2 c_s^2
-2\pi G \Sigma |k_r|\big]^{1/2}$ which
is the same as the dispersion relation
of Lin and Shu (1966) and Kalnajs (1965)
with the ``reduction factor'' ${\cal F}=1$.

Consider now equation (37)
for general $(k_r,k_\phi)$ with
$|{\bf k}|$ increasing but with
$\tan(\alpha)\equiv |k_\phi/k_r|=$const.
The
argument of the inner square root
in equation (37),
$(\kappa_r^2+{\bf k}^2u^2)^2
-8k_\phi^2 u^2\Omega r \Omega^\prime$,
has a minimum as a function of $|{\bf k}|$
at $|{\bf k}|= k_{TS}$, the same as for
$k_\phi=0$.
If the minimum of this argument is less
than zero, one of the mode frequencies
has a positive imaginary part (as well as
a real part for $\omega - m\Omega$).
This condition for instability can be
expressed as
\begin{equation}
{\Sigma_{crit} \over \Sigma_{TS}} > \left\{
1+{2qS^2 \over 2-q}-2\left[{qS^2 \over 2-q}
+{ q^2S^4 \over (2-q)^2}\right]^{1/2}\right\}^{1/2},
\end{equation}
where $S\equiv \sin(\alpha) =|k_\phi|/|{\bf k}|$
and $q \equiv -d\ln\Omega/d\ln r$.
For $S=1$ this gives
$\Sigma_{crit}/\Sigma_{TS} > 0.268$ (or
$Q<3.73$) for $q=3/2$
and
$\Sigma_{crit}/\Sigma_{TS} > 0.414$
(or $Q<2.42$) for $q=1$.
Thus, the non-axisymmetric modes
are unstable at larger $Q$ than
the axisymmetric modes.

For $|{\bf k}|>2k_{TS}$, we have
${\bf k}^2 u^2 >0$ so that the argument of
the inner square root in equation (37) is
greater than zero.
As a result, the outer square root in
equation (37) for $\pm_B=-1$ is
purely imaginary if $q>0$.
Thus the mode with $\pm_A=+1$
is unstable for $q>0$.

Figure 2 shows the four mode
frequencies as a function of $k_\phi$
for a case with
$\Sigma_{crit}<\Sigma < \Sigma_{TS}$.

\begin{figure}\begin{center}
\includegraphics[width=5in]{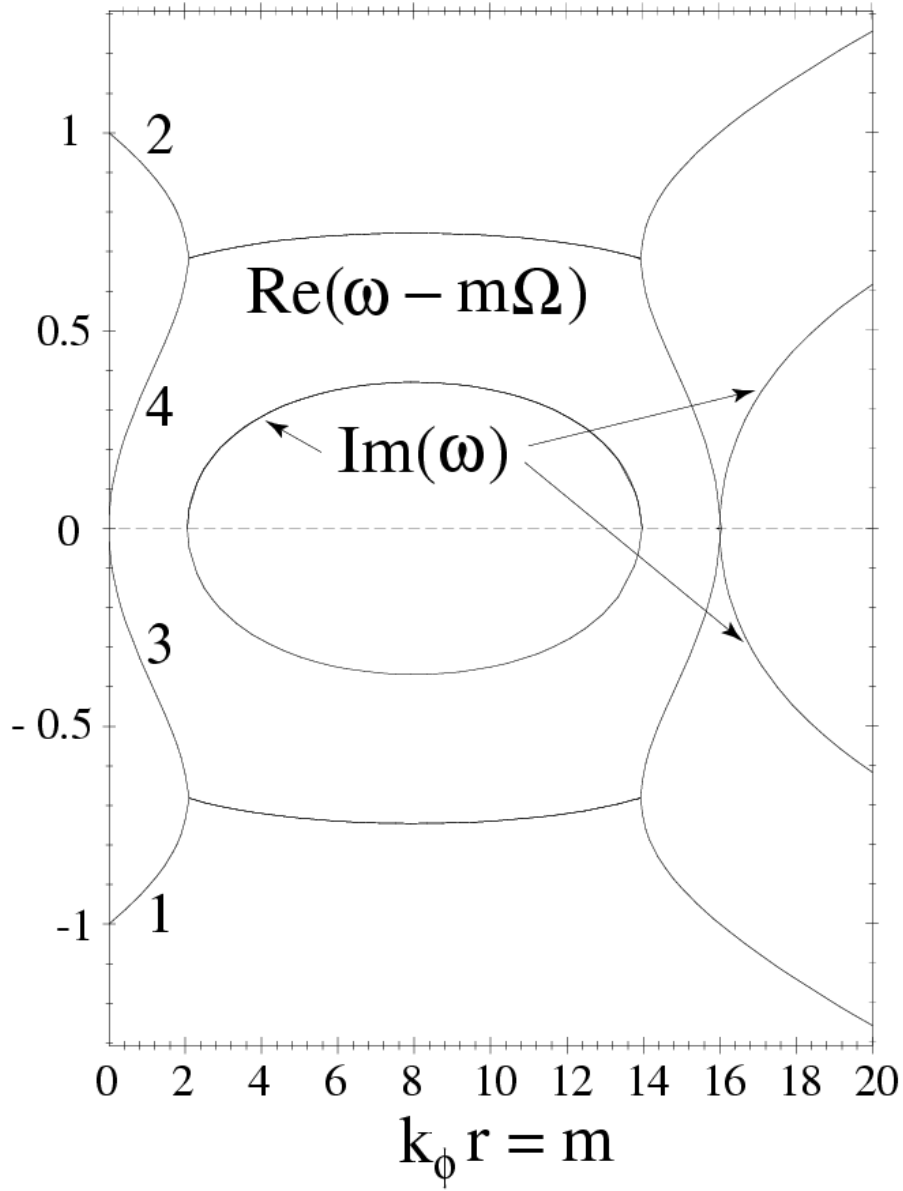}
\caption{Real and imaginary parts of
the four mode frequencies
(labeled $1,..,4$) normalized
by $\Omega$ for $k_\phi^2 \gg k_r^2$,
$c_s/(r\Omega)=0.05$,
and $\Sigma = 0.4\Sigma_{TS}$,
where $\Sigma_{TS}$ is the surface density
at which the Toomre-Safronov instability sets
in for axisymmetric perturbations.
We have taken
$q=3/2$ which
corresponds to a Keplerian disk.
For these conditions $rk_{TS}=8$.
}
\label{Figure2}
\end{center}
\end{figure}

\subsection{Shearing of Unstable Modes with
$k_\phi r \gg 1$}

For an unstable wave with
$k_\phi r = m \gg 1$,
the geometrical optics relations for a
wave-packet give
\begin{equation}
{d k_r \over dt} =- {\partial \omega \over \partial r}
\approx q k_\phi \Omega~,\quad
{d(r k_\phi) \over dt} =
-{\partial \omega \over \partial \phi}=0~,
\end{equation}
so that $k_r \approx q k_\phi \Omega t+$const.,
and $rk_\phi=$const.
Thus, a wave-packet initially with
long radial wavelength, with $|k_r|h \ll 1$,
is rapidly sheared by the differential
rotation of the disk if $q>0$.
In a time of order $t_{shear}=
(q k_\phi h \Omega)^{-1}$
the wave evolves into a very short radial
wavelength, $|k_r|h \gtrsim 1$, trailing spiral
wave.
This is discussed in detail by Goldreich
and Lynden-Bell (1965).
The maximum amplification a wave can have
(without reflections) is
$A=\exp(\int dt \omega_i) \sim
\exp[{\rm max}(\omega_i) t_{shear}]$,
where $\omega_i ={\cal I}m(\omega)$ is
the growth rate.
For the above-mentioned unstable
modes for $\Sigma_{crit}<\Sigma <
\Sigma_{TS}$, which have $k_\phi r \gg 1$,
we find $A \sim
\exp(1)$, which is probably
not significant in the absence
of reflections or
trapping of the waves in the vicinity
of a bump in the disk as a function of $r$
(Lovelace {\it et al.} 1999a; Li {\it et al.} 2000).

\section{Negative Energy Modes}

We consider the case of
a low mass disk around a Schwarzschild
black hole  where the inner radius of the
disk is at $r_i =3r_S$ with
$r_S \equiv 2GM/c^2$  the
Schwarzschild radius.
We adopt the pseudo-Newtonian
potential $\Phi = - GM/(r-r_S)$
(Paczy\'nski \& Wiita 1980;
Kato {\it et al.} 1998, ch. 2).
For this potential,
\begin{eqnarray}
\Omega=\left({GM \over r(r-r_S)^2}\right)^{1/2},\quad
{\kappa_r \over \Omega}=
\left({r-3r_S \over r-r_S}\right)^{1/2},
\nonumber\\
~~q \equiv -{d\ln\Omega \over d\ln r}
={3 \over 2}~{r-r_S/3 \over r-r_S }~.
\end{eqnarray}
Figure 3 shows the radial variations of
these quantities near the inner edge of
the disk. In the following the mode
frequency $\omega$
is measured in units of $\Omega_S\equiv
\sqrt{GM/r_S^3}$, while $\Omega$ is
in units of ${\rm s}^{-1}$.
Radial distance $R$ and $1/k_r$
are measured
in units of $r_S$, while ${\cal E}$ and
$\Delta$
are in units of length.
Note that $\Omega_S \approx
7.16\times 10^4 (M_\odot/M){\rm s}^{-1}$
and $r_S \approx 2.96\times
10^5(M/M_\odot){\rm cm}$.

\begin{figure}\begin{center}
\includegraphics[width=5in]{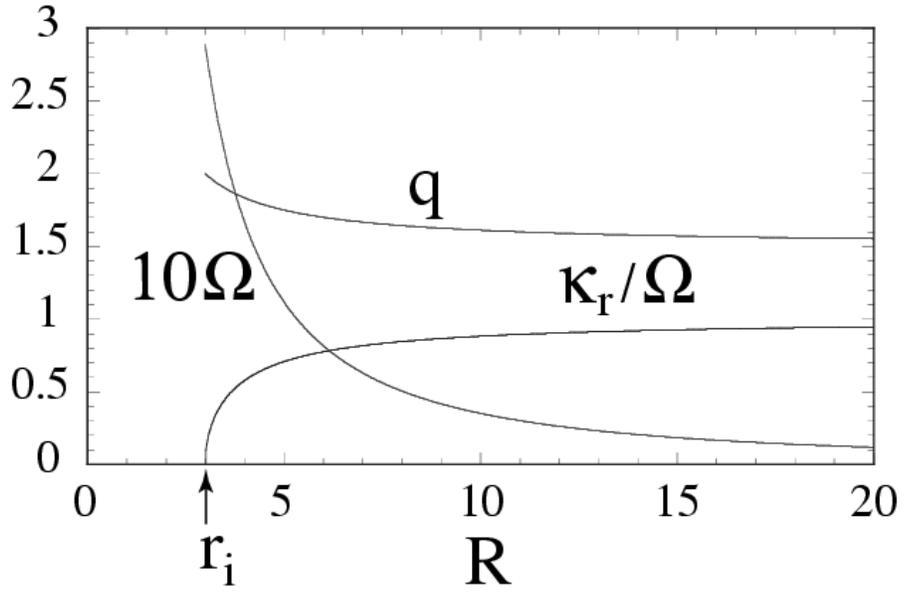}
\caption{Radial variations of equilibrium disk
quantities of equation (41) with $R\equiv r/r_S$
and $\Omega$ measured in units of $\Omega_S \equiv
(GM/r_S^3)^{1/2}$.
}
\label{Figure2}
\end{center}
\end{figure}

\subsection{One-Armed Spiral Waves {\rm (}$m=1${\rm )}}

Figure 4 shows the radial wavenumber
dependence of the low-frequency mode $\omega_1$
and the corresponding mode energy
(per unit
area) $dE/dA$ (from equation (31)) for a sample case
for $k_r \geq 0$ which corresponds
to a {\it trailing}
spiral wave.
(The $1-$subscript on $\omega$ is
dropped in the following.)
The radial phase velocity is outward
($\omega/k_r >0$) and
the azimuthal phase velocity is
positive ($\omega r>0$).
[Note that both quantities plotted
in Figure 4 are
even functions of $k_r$, so that
the plot is the same for $k_r<0$
which corresponds to a leading spiral
wave.]
For $k_r$ less than
a certain value $k_{r0}$, the
wave frequency is positive while
the wave energy is {\it negative}.
Further, the energy flux-density
of the perturbation ${\cal F}_{Er}$
(from equation (32))
is positive (outward) for $k_r < k_{r0}$
and negative for $k_r>k_{r0}$.

\begin{figure}\begin{center}
\includegraphics[width=5in]{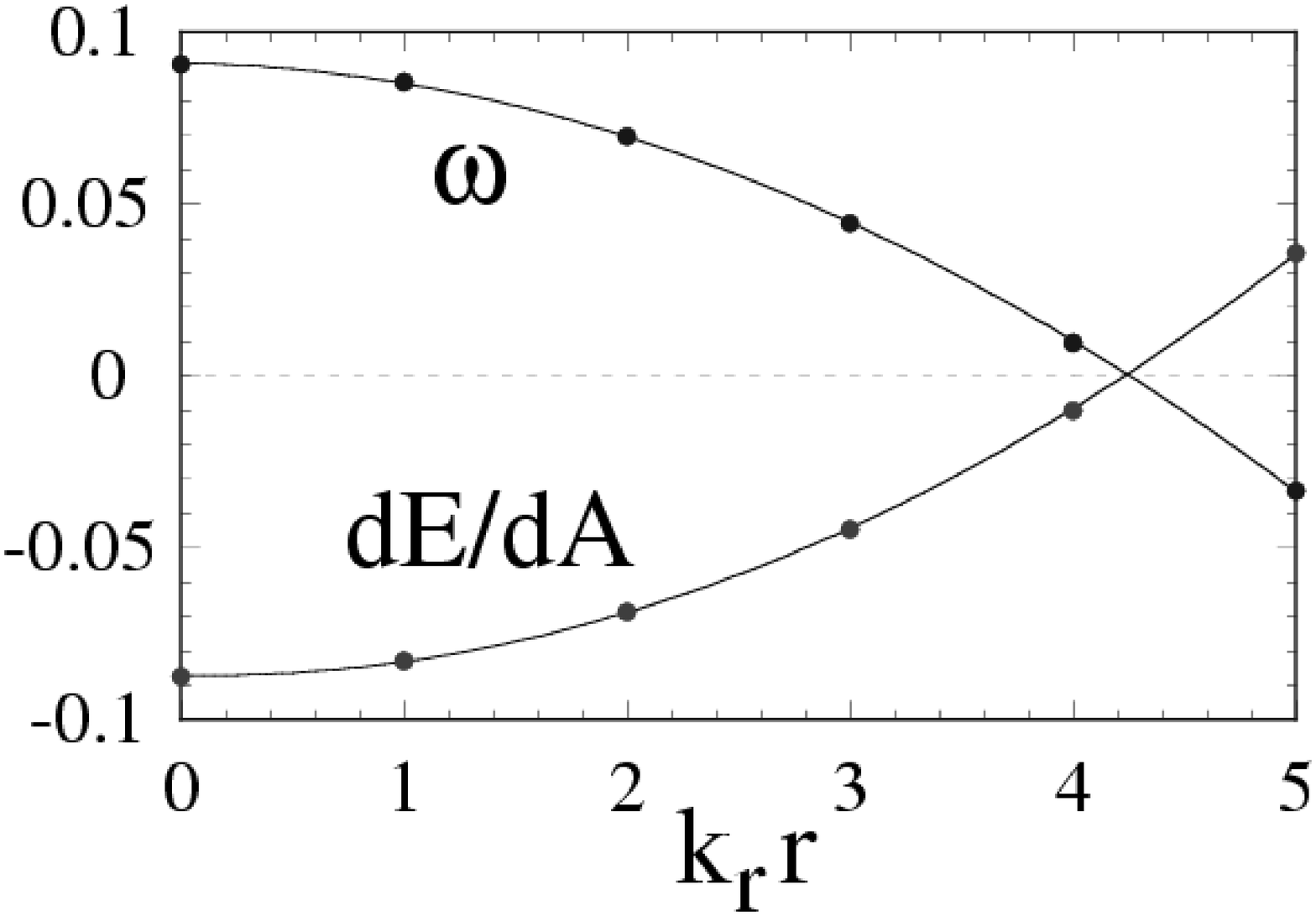}
\caption{
Dependence
of the low-frequency mode $\omega$
on radial wavenumber
from equation (37)
and the corresponding
wave energy (per unit area) ${dE/dA}$
from equation (31).
For $k_r<0$ note that
both curves are even functions of
$k_r$.
Here, $\omega$ is normalized by
$\Omega_S$, and $dE/dA$
by $\Sigma \Omega^2 |{\cal E}|^2/2$.
For this plot $R=10$ and $c_s/(r\Omega)=0.1$.}
\label{Figure4}
\end{center}
\end{figure}

From Figure 4 the group
velocity $v_{gr} = \partial \omega/\partial
k_r$ is negative for $k_r >0$.
Thus, the sign
of ${\cal F}_{Er}/(dE/dA) < 0$ is the same
as that of $v_{gr}$ for
$k_r {\buildrel < \over >}
k_{r0}$ as it should be.
Note that the radial phase and group
velocities have opposite signs.

The canonical angular
momentum (per unit area)
${\cal P}_\phi$ (from equation 33)
is negative and
the flux of angular momentum
${\cal F}_{{\cal P}r}$ (from equation 34)
is positive for the
low frequency mode for
$k_r {\buildrel < \over >}k_{r0}$.
Thus ${\cal F}_{{\cal P}r}/{\cal P}_\phi <0$
has the same sign as $v_{gr}$.

The signs of the different
quantities are summarized
in Table 1 for the trailing waves
($k_r>0$).
For the leading
waves, $k_r \rightarrow -k_r$, the signs
of $dE/dA$ and ${\cal P}_\phi$ do not change,
but the signs of the fluxes ${\cal F}_{Er}$
and ${\cal F}_{{\cal P}r}$ reverse as
does the sign of the group velocity $v_{gr}$.

\begin{table*}
\caption{Physical Quantities
for the Low-Frequency Mode}
\begin{center}
\begin{tabular}{ccccccc}
\hline
\hline
\rule[-4pt]{0pt}{20pt}
{} & {$\omega$} &
{$dE/dA$} & {${\cal F}_{Er}$} &
{$v_{gr}=\partial \omega / \partial k_r$}
&{${\cal P}_\phi$} &
{${\cal F}_{{\cal P}r}$}\\
\hline
{$0<k_r<k_{r0}$} & {$+$} & {$-$} &
{$+$} & {$-$} & {$-$} & {$+$} \\
\hline
{$k_{r}>k_{r0}$} & {$-$} &
{$+$} & {$-$} & {$-$} & {$-$} & {$+$}\\
\hline
\end{tabular}
\end{center}
\end{table*}

In contrast, for the high-frequency mode
$\omega_2$, the wave energy is {\it positive}
independent of $k_r$.
Also, the angular
momentum ${\cal P}_\phi$ is positive.

For the low-frequency
mode the approximate dependence is
\begin{equation}
\omega = \omega_0 \left(
1- {(k_r r)^2 \over (k_r r)_0^2} \right)~.
\end{equation}
Figure 5 shows the $R-$dependences
of $\omega_0$
and $(k_r r)_0$ for
$c_s/(r\Omega)=0.1$.
For this case
$\omega_0 \approx C_\omega/R^a$ with
$C_\omega = 4.71$, $a=3.16$
and $(k_r r)_0 \approx C_k/R^b$
with $C_k=21.1$, $b=0.702$.
The frequency is of course independent
of $R$ so that
\begin{equation}
k_r^2={C_k^2 \over R^{2+2b}}
\left(1 -{\omega R^a \over C_\omega}\right)~,
\end{equation}
where the allowed region $k_r^2 \geq 0$
extends from the inner radius of the
disk $R=R_i=3$ out to the turning point
$R_{turn}=(C_\omega/\omega)^{1/a}>3$.

Because $dE/dA \approx
- \omega \Sigma \Omega^2 |{\cal E}|^2/2$,
we have
\begin{equation}
E \approx -\pi r_S^2
~\omega \int_{R_i}^\infty
R dR ~\Sigma \Omega^2 |{\cal E}|^2~,
\end{equation}
which is negative in that $\omega>0$
for the `bound states'.
Also, we find ${\cal P}_\phi
\approx - \Sigma\Omega |{\cal E}|^2$
so that the total canonical angular
momentum of the disk perturbations is
\begin{equation}
L_z \approx -\pi r_S^2 \int_{R_i}^\infty
RdR~\Sigma \Omega |{\cal E}|^2~.
\end{equation}
We discuss the mode energy and
angular momentum further in \S 5.3.

\begin{figure}\begin{center}
\includegraphics[width=5in]{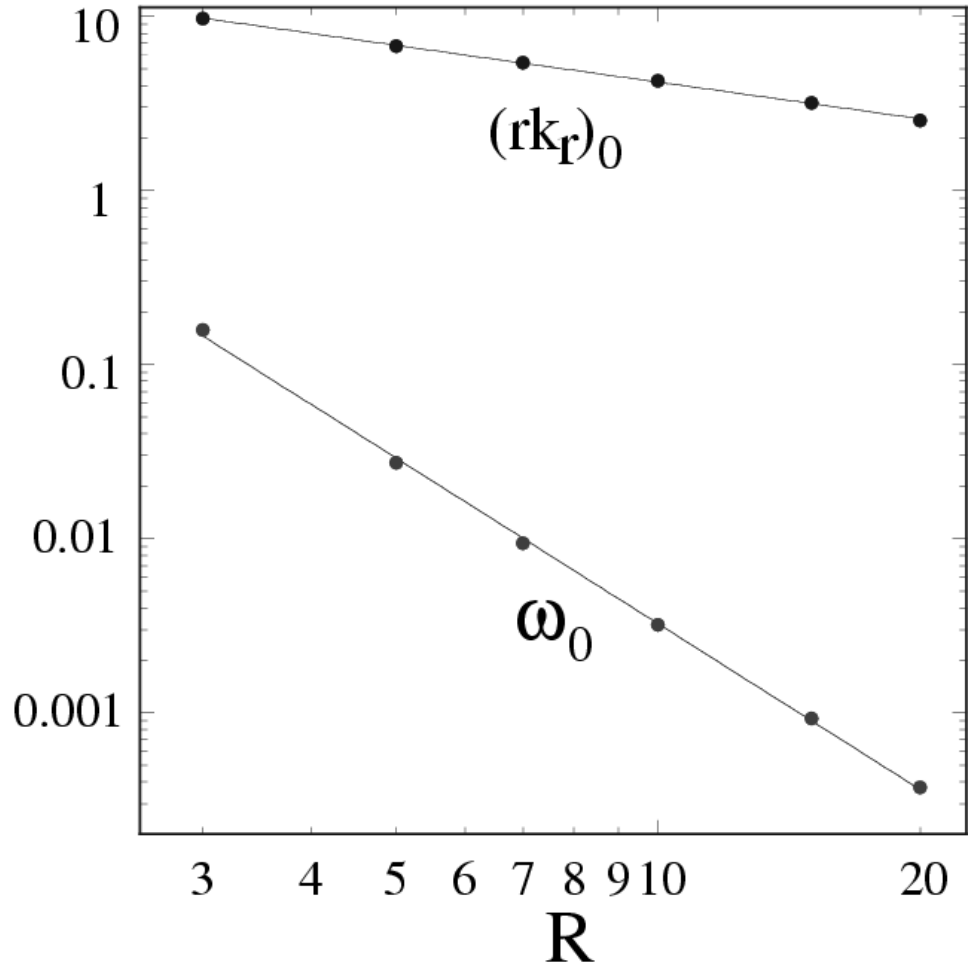}
\caption{Radial dependences of the two
`constants' in equation (42) for the
case where $c_s/(r\Omega)=0.1$.
The straight lines correspond to
least squares fits through
the shown points which give $(k_r r)_0
=21.1/R^{0.702}$ and $\omega_0=4.71/R^{3.16}$.
}
\label{Figure5}
\end{center}
\end{figure}

In the limit
$\omega \rightarrow 0$, the
trailing spiral wave for example has the
form $R=R_i/(1+b3^b \phi/C_k)^{1/b}$
for $\phi\leq 0$.

\subsection{Standing Waves}

A standing wave solution
for say ${\cal E}$ can
be written as
\begin{eqnarray}
{\cal E}&=& {\cal E}_+
+ {\cal E}_- ~,
\nonumber \\
&=& A |k_r|^{-1/2}\exp(i\Psi)+
B |k_r|^{-1/2}\exp(-i\Psi)~,
\end{eqnarray}
where the ${\cal E}_+$ term corresponds
to a trailing spiral wave and the
${\cal E}_-$ term a leading wave.
Also,
\begin{equation}
\Psi(R,\omega) \equiv \int_{R_i}^{R}
dR^\prime ~ k_r(R^\prime,\omega)
\end{equation}
is the phase with the $\omega$
dependence made explicit.
An expression for $\Delta$ in
terms of ${\cal E}$ can be derived
from equations (35) and (36).

The specific nature of the modes
clearly requires a definite boundary
condition at the inner edge of the
disk.
The present model, which
neglects $v_r$, does not allow
a full treatment of the inner
radius of the
disk where the radial inflow
may be transonic (see for example
Kato {\it et al.} 1998, ch. 9).
Instead, we first consider an inner
disk boundary condition appropriate
for the Shakura and Sunyaev (1973)
model where for $r \rightarrow r_i$,
$\Sigma \rightarrow
\infty$; namely,
${\cal E}(r_i=3r_S)=0$, which corresponds
to perfect reflection.
For this boundary condition,
and the usual WKBJ matching at the outer
turning point $R_{turn}$, we obtain the
Bohr-Sommerfeld quantization condition
\begin{equation}
\Psi(R_{turn},\omega) =\left(n +
{3 \over 4}\right)\pi~,\quad n=0,1,2,..~,
\end{equation}
which determines the allowed $\omega$
values.
The $3/4$ in equation (48)
(rather than $1/2$) is due to the
boundary condition at $R_i$.
For the mentioned dependences
of $\omega_0$ and $(k_r r)_0$, and
$c_s/(r\Omega)=$ const, there are
a finite number of modes $n$
with the lowest frequency corresponding
to the largest $n$ value.

\begin{figure}\begin{center}
\includegraphics[width=5in]{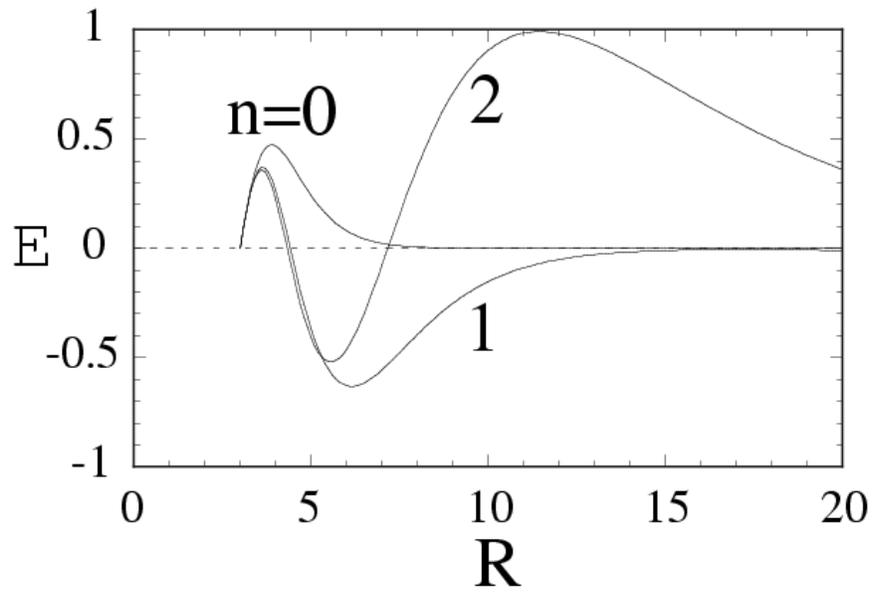}
\caption{
Wave function profiles ${\cal E}_n(R)$
for the three low-frequency
standing wave modes for
the case $c_s/(r\Omega)=0.1$ obtained
by solving equation (49) using a shooting
method.
The normalization of the
different wavefunctions results
from the boundary
condition $(d{\cal E}_n/dR)_i=1$.
}
\label{Figure6}
\end{center}
\end{figure}

Instead of using the approximate
equation (48),
we solve the wave equation
for the perturbation amplitude,
\begin{equation}
{d^2 {\cal E} \over dR^2} =
- k_r^2(R,\omega)~ {\cal E}~.
\end{equation}
This is done using a shooting method:
We guess $\omega$, integrate outward from
$R=R_i=3$, and require
${\cal E}(R\rightarrow \infty) \rightarrow 0$.
This requirement determines the allowed
frequency values.
Figure 6 shows the behavior of the
wave functions for a sample case
with $c_s/(r\Omega)=0.1$.
The corresponding frequencies
in units of $\Omega_S$
are $\omega_n \approx
3.75\times 10^{-2}$, $7.74\times 10^{-3}$, and
$8.69 \times 10^{-4}$ for $n=0,1,2$,
respectively.
For a $M=10M_\odot$ black hole,
these frequencies in Hz are: $268,~55.4$,
and $6.22$.
These frequencies
agree approximately
with those obtained from equation (48).
This equation implies
that there are only
three modes (or bound states) for
$c_s/(r\Omega)=0.1$.

\begin{figure}\begin{center}
\includegraphics[width=5in]{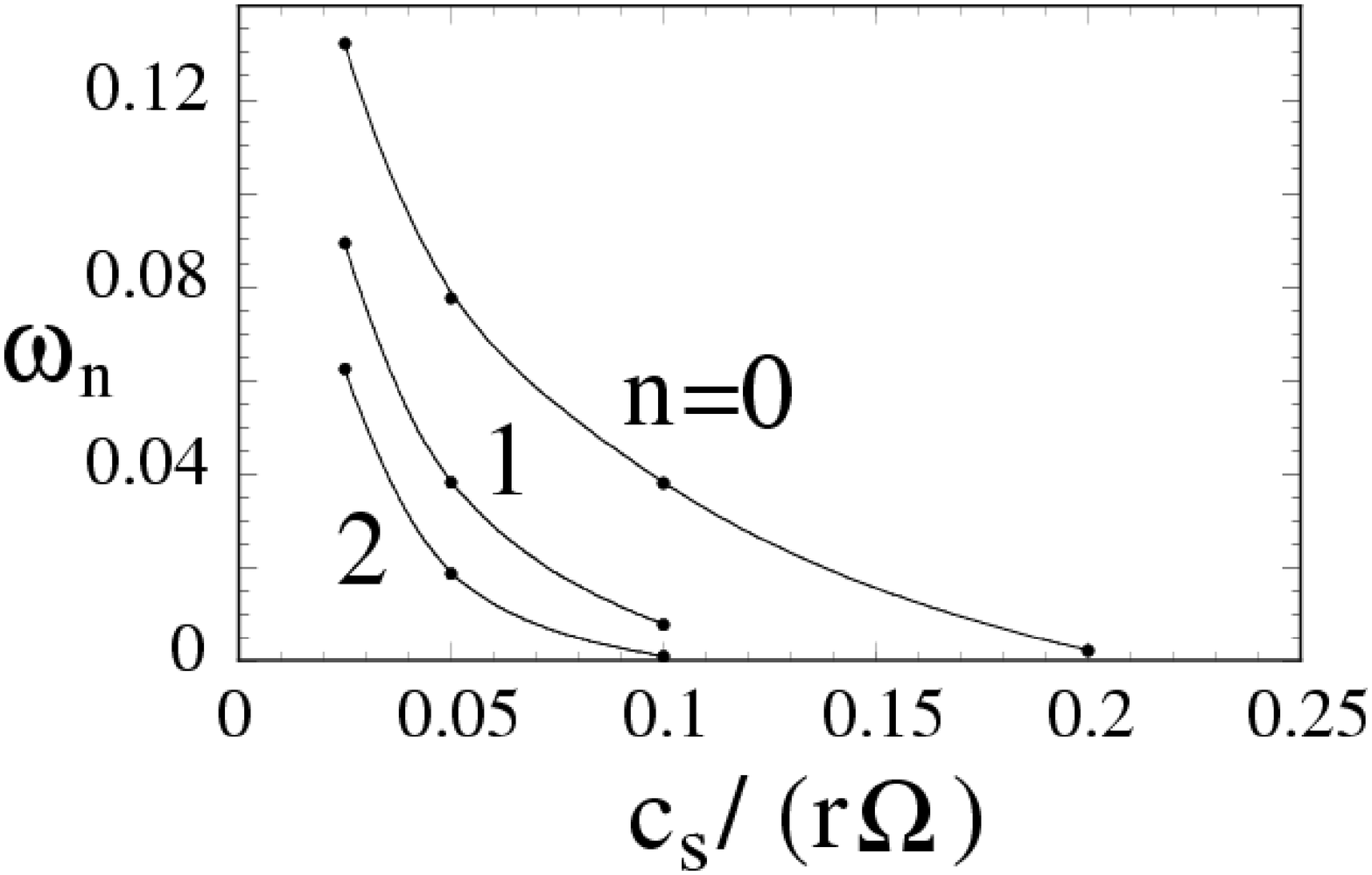}
\caption{Dependence of mode
frequencies $\omega_n$ in units
of $\Omega_S$ on the disk temperature
as measured by $c_s/(r\Omega)$.
The curves are smooth fits through the
shown points.
}
\label{Figure 7}
\end{center}
\end{figure}

Figure 7 shows the variation of the
mode frequencies with the disk thermal
speed $c_s/(r\Omega)$.

\begin{figure}\begin{center}
\includegraphics[width=5in]{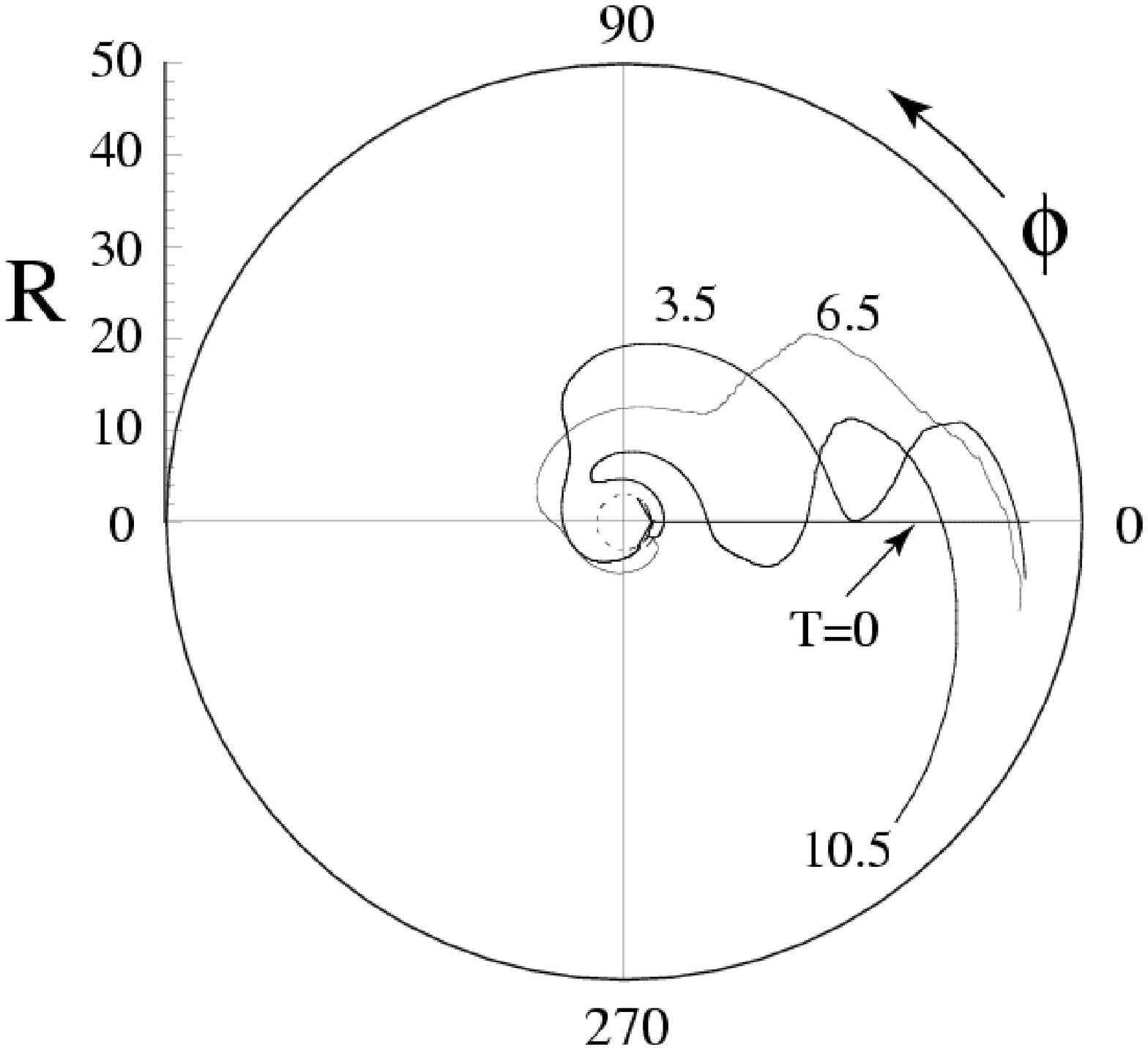}
\caption{
Time dependence of the line-of-nodes
angle $\varphi(R,T)$, which is such that
${\cal E}=|{\cal E}(R,T)|\exp[-i\varphi(R,T)]$,
obtained by
solving the Schr\"odinger equation (51)
for an initial wavefunction ${\cal E}(R,T=0)
={\rm const}(R-3)\exp(-R/4)$ which is
{\it not} an eigenfunction.
The boundary conditions
were ${\cal E}(R_i)=0=
{\cal E}(R_{max})$ and a nonuniform
mesh $R_j=R_iQ^j,~j=1,..,300$ with $Q=1.009$
so that $R_{max}\approx 44.1$.
The time-dependent evolution
represents a superposition of the
three eigenfunctions shown in
Figure 6 with the average of
the energy $<\tilde E> =- 8.09 \times
10^{-3}w$.
The plot shows the reflection of
a trailing spiral wave ($T \lesssim 125$)
from the inner
disk radius which gives rise to a leading
spiral wave ($T \gtrsim 125$) for
$3\leq R \lesssim 5.6$.}
\label{Figure 8}
\end{center}
\end{figure}

With regard to the energy of the
standing wave modes, notice that
we now have in place of equations
(28) separate equations for ${\cal E}_+$,
${\cal E}_-$, $\Delta_+$, and $\Delta_-$.
Thus the Lagrangian density becomes
${\cal L}={\cal L}_+({\cal E}_+,..)
+{\cal L}_-({\cal E}_-,..)$, and
the Hamiltonian ${\cal H}=
{\cal H}_+ + {\cal H}_-$.
The energy of the trailing and
leading components are equal and
negative.
Thus the total energy $E = E_{+}+E_{-}$
is negative.
From equation (32) the total energy
flux is
\begin{eqnarray}
{\cal F}_{Er} &=&
\omega \Sigma c_s^2~
{\cal I} m \left({\cal E}_+^*
{\partial {\cal E}_+ \over \partial R}
+{\cal E}_-^* {\partial {\cal E}_-
\over \partial R}\right)~,
\nonumber\\
&= &
\omega \Sigma c_s^2~
{\cal I} m \left({\cal E}^*
{\partial {\cal E} \over \partial R}
\right)~,
\nonumber\\
&=&\omega \Sigma c_s^2 |k_r|
\left(|{\cal E}_+|^2 -|{\cal E}_-|^2 \right)~.
\end{eqnarray}
For the considered boundary
condition at the inner disk radius
$R=3$, ${\cal F}_{Er}=0$.
At large distances, ${\cal F}_{Er}=0$
because ${\cal E}(R\rightarrow \infty) =0$.
Thus from equation (21)
we have $d E/dt = 2\pi (r{\cal F}_{Er})_i
-2\pi(r{\cal F}_{Er})_\infty
=0$, where the $i-$subscript indicates
evaluation at $R=3$.
Similar conclusions apply to the
canonical angular momentum where
${\cal F}_{{\cal P}r} =
\Sigma c_s^2 |k_r|
(|{\cal E}_+|^2 -|{\cal E}_-|^2)$.

\subsection{Time-Dependent Solutions}

Equation (49) can
be written in the
form of a Schr{\"o}dinger equation
\begin{equation}
{w \over i}
{\partial {\cal E} \over \partial T}=
\left[-{\partial^2 \over \partial R^2}
+U(R) \right] {\cal E}~,
\end{equation}
where $T \equiv \Omega_S t$,
$U(R) \equiv - C_k^2/R^{2+2b}$ and
$w \equiv (C_k^2/C_\omega)/R^{2+2b-a}$.
The $R-$dependence of
$w$ is rather weak ($\sim R^{-0.24}$ for
$c_s/(r\Omega)=0.1$) and for this
reason we make the simplification
that $w=w(R=R_i)$.
The sign of the time derivative is
opposite that of Schr\"odinger equation
because the energy is negative for
$\omega >0$.

For the mentioned bound states,
$\tilde{E}_n =-w\,\omega_n$,
where the tilde indicates a different
normalization of the energy
compared with equation (44).
For the unbound states or continuum
modes we have $\omega_k <0$
and $\tilde{E}_k= -w\, \omega_k>0$.
For the continuum trailing
waves ($k_r>0$) both the radial
phase and group velocities are negative,
whereas for the leading waves ($k_r<0$)
both are positive.

Figure 8 shows polar plot (see
Briggs 1990; LZKH)
of the time-dependence of the
mode for an assumed initial perturbation
${\cal E}(R,T=0)={\rm const}(R-R_i)\exp(-R/4)$
which is neither leading nor trailing.
   The mode  evolves to
a predominantly {\it trailing spiral
wave}. Trailing spiral one-armed spiral
waves were found under different
conditions in our
earlier study of galactic disks (LZKH).

\section{Conclusions}

The perturbations of thin disks
have been studied in many works applied
to understanding the waves or
modes in galactic disks
and in accretion disks.
This work has derived the linearized
equations of motion, the
Lagrangian density, the Hamiltonian
density, and the
canonical angular momentum density
for general perturbations [$\propto
\exp(im\phi)$ with $m=0,\pm 1,..$]
of a geometrically thin
self-gravitating, homentropic fluid disk
including the pressure.
The theory was applied
to the ``eccentric,'' $m=\pm 1$ perturbations
of a geometrically thin Keplerian disk.
The $m=1$ modes may be at low frequencies
compared to the Keplerian frequency.
Furthermore, the modes
can have negative energy and
negative angular momentum.
Their propagation  can  remove
angular momentum from the inner region of a disk
and thus act to enhance the accretion of matter.
Dependent on the radial boundary conditions,
there are discrete, low-frequency, negative
energy $m=1$ modes.
Time-dependent perturbations tend to
give rise to predominantly trailing spiral waves.
   A future direction for extending
this paper include the treatment
of the non-linear evolution following
the approach of
 Shu, Yuan, and
Lissauer (1985), Heemskerk,
Papaloizou, and Savonije (1992),
and Lee and Goodman 1999).

\bigskip

We thank M.P. Haynes,
D.A. Kornreich, N.F. Comins, and M.M. Romanova for
stimulating discussions on eccentric
perturbations of disks.
This work was supported in part
by NASA grants NAGS-9047, NAGS
9735, by NSF grant AST-9986936,
and by CRDF grant KP-2555-AL-03.


\begin{thebibliography}{}

\bibitem{} Adams, F. C., Ruden, S. P.,
\& Shu, F. H. 1989, ApJ, 347, 959

\bibitem{} Andersson, N. 1998, ApJ, 502, 708

\bibitem{} Baldwin, J.E., Lynden-Bell, D., \& Sancisi, R.
1980, MNRAS, 193, 313

\bibitem{} Binney, J., \& Tremaine, S. 1987,
{\it Galactic Dynamics}
(Princeton: Princeton University Press)

\bibitem{} Briggs, F.H. 1990, ApJ, 352, 15

\bibitem{} Coppi, B., Rosenbluth, M.N.,
\& Sudan, R.N. 1969, Ann. of Physics, 55, 207

\bibitem{} Fridman, A.M., Boyarchuk, A.A., Bisikalo, D.V.,
Kuznetsov, O.A., Khorushii, O.V., Torgashin, Yu.M., \&
Kilpio, A.A. 2003, Physics Letters A, 317, 181

\bibitem{} Frieman, E., \& Rotenberg, M. 1960,
Rev. Mod. Phys., 32, 898

\bibitem{} Goldreich, P., \& Lynden-Bell, D. 1965,
MNRAS, 130, 125

\bibitem{} Goldstein, H. 1950, {\it Classical
Mechanics} (Addison-Wesley: Cambridge, Massachusetts)

\bibitem{} Goodman, J., Narayan, R., \&
Goldreich, P. 1987, MNRAS, 225, 695

\bibitem{}Heemskerk, M. H. M., Papaloizou, J. C.,
\& Savonije, G. J. 1992, A\&A, 260, 161

\bibitem{} Kalnajs, A.J. 1965, Ph.D. thesis, Harvard
University

\bibitem{} Kato, S., Fukue, J., \& Mineshige, S. 1998,
{\it Black-Hole Accretion Disks} (Kyoto University Press:
Kyoto, Japan)

\bibitem{} Kornreich, D.A., Haynes, M.P., \&
Lovelace, R.V.E. 1998, AJ, 116, 2154

\bibitem{} Kornreich, D.A., Lovelace, R.V.E.,
\& Haynes, M.P. 2002, ApJ, 580, 705

\bibitem{} Lee, E., \& Goodman, J. 1999, MNRAS, 308, 984

\bibitem{} Li, H., Finn, J.M., Lovelace, R.V.E., \&
Colgate, S.A. 2000, ApJ, 533, 1023

\bibitem{} Lin, C.C., \& Shu, F.H. 1966,
Proc. Nat. Acad. Sci., 55, 229

\bibitem{} Lovelace, R.V.E. 1998, A\&A, 338, 819

\bibitem{} Lovelace, R.V.E., Li, H., Colgate, S.A., \&
Nelson, A.F. 1999a, ApJ, 513, 805


\bibitem{} Lovelace, R.V.E., Zhang, L., Kornreich, D.A.,
\& Haynes, M.P. 1999b, ApJ, 524, 634


\bibitem{} Lynden-Bell, \& Ostriker, J.P. 1967,
MNRAS, 136, 293

\bibitem{} Nowak, M.A., \& Wagoner, R.V. 1991, ApJ, 378, 656

\bibitem{} Nowak, M.A., \& Wagoner, R.V. 1992, ApJ, 393, 697

\bibitem{} Paczy\'nski, B., \& Wiita, P.J. 1980,
A\&A, 88, 23

\bibitem{} Rix, H.-W., \& Zaritsky, D. 1995,
ApJ, 447, 82

\bibitem{} Safronov, V.S. 1960, Ann. Astrophys., 23, 982

\bibitem{} Shakura, N.I., \& Sunyaev, R.A. 1973,
A\&A, 24, 337

\bibitem{} Schenk, A. K., Arras, P., Flanagan, \'{E}. \'{E}., Teukolsky, S. A.,
\& Wasserman, I. 2002, Phys. Rev. D, 65, 024001

\bibitem{} Shu, F. H., Yuan, C.,
\& Lissauer, J. J. 1985, ApJ, 291, 356

\bibitem{} Shu, F. H., Tremaine, S., Adams, F. C.,
\& Ruden, S. P. 1990, ApJ, 358, 495

\bibitem{} Toomre, A. 1964, ApJ, 139, 1217

\bibitem{} Tremaine, S. 2001, AJ, 121, 1776

\bibitem{} van der Klis, M. 2000, Ann. Rev. of Astron.
and Astrophysics, 38, 717

\bibitem{} Zeltwanger, T., Comins, N.F., \& Lovelace, R.V.E. 2000,
ApJ, 543, 669








\end{thebibliography}
\end{document}